\documentclass[twocolumn,PRD,showpacs,nofootinbib,amsmath,amssymb]{revtex4}

\usepackage{epsfig}

\begin{document}
\title{Power spectrum nulls due to non-standard inflationary evolution}
\author{Gaurav Goswami\footnote{gaurav@iucaa.ernet.in} and Tarun Souradeep\footnote{tarun@iucaa.ernet.in} }
\address{IUCAA, Post Bag 4, Ganeshkhind, Pune-411007, India}
\date{\today}

\begin{abstract}
The simplest models of inflation based on slow roll produce nearly
scale invariant primordial power spectra (PPS). But there are also numerous models that
predict radically broken scale invariant PPS.  In particular, markedly
cuspy dips in the PPS correspond to nulls where the perturbation amplitude, hence PPS, goes
through a zero at a specific wavenumber. Near this wavenumber, 
the true quantum nature of the generation mechanism of the primordial fluctuations may be revealed. 
Naively these features may appear to
arise from fine tuned initial conditions. However, we show that this
behavior arises under fairly generic set of conditions involving
super-Hubble scale evolution of perturbation modes during inflation.
We illustrate this with the well-studied examples of punctuated
inflation and the Starobinsky-break model.
\end{abstract}
\maketitle

The paradigm of cosmological inflation explains not only a set of
peculiarities, such as the flatness problem, the horizon problem,
etc., in the hot big bang model, but also the origin of the initial
metric perturbations that led to formation of the large scale
structure in the distribution of matter in the universe
\cite{inflation,fluctuations,infrev}.
The simplest models of inflation achieve this by assuming that at high
enough energy scales, the dynamics of the universe is as if it was
dominated by a single scalar (inflaton) field. The inevitable quantum
fluctuations seed the primordial metric perturbations.

\begin{figure}
\centerline{\psfig{figure=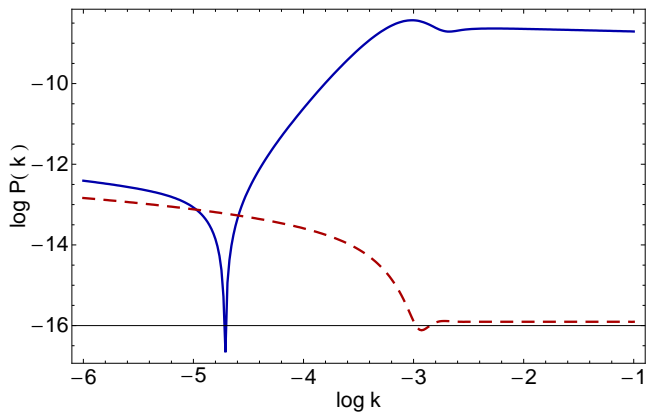, width = 8.5 cm, angle=0}}
\caption{\small\sf The scalar (thick, blue) and tensor (dashed, red) PPS in Punctuated inflation model, 
an example of a sandwich model.
Notice the sharp spike in scalar PPS and tensor power overtaking scalar power for for a range of modes 
near it.}
\label{PI_PPS}
\end{figure}

The primordial power spectrum (PPS) is connected to observed angular
power spectrum ($C_l$s) of the temperature fluctuations in the CMB sky
through the radiative transport kernel. Alternatively, for a given
cosmology (which determines the transport kernel), the primordial power
spectrum can be deconvolved from the observed
$C_l$s~\cite{ppsfeatures}. These results seem to indicate that the
PPS may have features, e.g., a sharp infrared
cutoff on the horizon scale, a bump (i.e. a localized excess just
above the cut off) and a ringing (i.e. a damped oscillatory feature
after the infrared break). While the statistical significance of such
a feature is still being assessed~\cite{armantarundec09}, this led to
a lot of activity in building up models of inflation that can give
large and peculiar features in primordial power spectrum (see
\cite{PI1} and references therein, along with 
\cite{Hodges1990,double-inf,Kof-linde,
Staro1992,cusps,nonTevol,nonTevol2,Leach1}). Many such models
tend to assume very special initial conditions at the beginning of
inflation. In contrast, others postulate quite fine tuned values of
the parameters of the Lagrangian of the inflaton at tree level to
produce features in the scalar PPS 
\cite{PI1,cusps,Hodges1990}).
In many such scenarios, the scalar PPS has cuspy dips~
 \cite{PI1,cusps,Leach1,nonTevol2} (also, see Fig. \ref{PI_PPS}) that
actually correspond to a null in the PPS i.e. precisely zero scalar power at some wave number . 
Also for a
range of modes near such a feature, the tensor power overtakes scalar
power \cite{PI2}. Scalar PPS with cuspy dips turn up in many models of inflation with different forms of potential 
(such as false vacuum inflation model with quartic potential \cite{Leach1}, double-well potential \cite{nonTevol2}, 
Coleman-Weinberg potential \cite{nonTevol2} etc) and also in other ways (such as in dissipative models of inflation, see \cite{HP}). 
It is seen that when one tries to produce 
an enhanced power on some scale 
(such as when considering models which try to enhance the production of primordial black holes), 
it is accompanied with a sharp drop in power leading to cusps in scalar PPS. 
Similarly, models that tend to produce low power in low multipoles of CMB anisotropies end up 
having sharp cusps. Thus, cusps in scalar PPS have been reported in the literature but their origin
 has not been satisfactorily understood. An exact null in scalar PPS can have interesting consequences such 
as on processed non-linear power spectrum. On the other hand, this null in the power spectrum is found by doing 
a classical computation, thus, for a small range of modes near the one having a zero, quantum effects can not be 
neglected \cite{quantum} and hence the null may not be present when the quantum effects are taken into account 
exposing the truly quantum nature of the generation mechanism of primordial perturbations. This motivates us to 
study the origin and conditions required for cuspy dips in scalar PPS.

In this paper we take a fresh look at the evolution of mode functions
of cosmological perturbations during inflation, a subject that is well studied (see \cite{pascal} 
for a recent treatment). 
We point out that
there are some key properties that the mode functions follow as they
evolve in the complex plane. We also realize that it is possible to cause
a particular kind of non standard evolution of modes. This kind of non standard 
evolution is connected closely to the existence of sharp cuspy dips in the scalar PPS.

We study the complex plane trajectory that the Fourier mode function
of perturbation variables such as the curvature perturbation on
hypersurfaces orthogonal to comoving worldlines, $R_k$ and the
Mukhanov-Sasaki variable $v_k=z R_k$ (where $z=a \dot{\phi} /H$).
The evolution of $v_k$ is given by

\begin{equation} \label {MSE}  
 {v_k}'' + \left( k^2 - \frac{z''}{z} \right)v_k = 0 \, .
\end{equation}

\begin{figure}[t]
\begin{center}
\epsfig{file=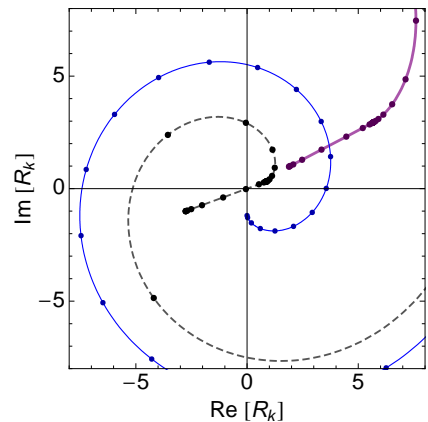,angle=0,width=5.5cm,height=5.5cm}
\end{center}
\caption[] {\small\sf The trajectories of mode function of $R_k$ in
complex plane. The dots on the trajectory are uniform in expansion e-folds,
crowding of dots indicates that
$R_k$ freezes. The thin trajectory corresponds to power law inflation which never crosses the origin. Notice that when $R_k$ freezes, the tangent vector to
its trajectory points radially in.  
The other two trajectories show the behavior of $R_k$ in the complex plane for two different modes in
Punctuated inflation model. The mode
denoted by continuous (thick) purple curve freezes once, then unfreezes, evolves radially in and freezes again
before crossing the origin. Dashed gray curve corresponds to a mode with larger $k^2$, it undergoes large radial
super-Hubble evolution, crosses the origin and then freezes undergoing super-Hubble enhancement of power compared with the other mode that undergoes suppression.
In between would lie a mode (not shown) that ends up right at the origin leading to a null in PPS.}
\label{stdR}
\end{figure}

Above equation shows that the mode function of $v$ goes along a circle
of radius $1/\sqrt{2 k}$ in clockwise sense (Bunch-Davies vacuum 
\footnote{It is important to note that except for the  exact shape of the trajectory, the general 
arguments that we have given do not depend on the choice of vacuum.}) in
the complex plane when the mode is well inside the Hubble radius
($k\gg aH$) and goes radially outwards, (with $v_k\propto z$) when the
mode is well outside the Hubble radius ($k\ll aH$), see e.g. \cite{nonTevol}. Correspondingly,
$R_k$ just spirals-in (along the curve with polar equation $r \sim \theta ^{\nu -1/2}$ for power law inflation) in 
extreme sub-Hubble regime, while it just
freezes to some value when the mode is in super-Hubble regime. It is important to
note that just prior to freezing, the tangent vector to the trajectory
of $R_k$ points radially inward since, $z$ is negative, 
see the thin trajectory in Fig. \ref{stdR}. We
will refer to this as standard evolution of the mode function of
$R_k$. It is easy to confirm that in the simplest case of power law
inflation, the trajectory of $R_k$ in the complex plane would never
cross the origin.

It is a well known fact that in a universe dominated by a single scalar
field, $R_k$ always freezes on super Hubble scales.
This means that the amplitude as well as the phase of $R_k$ freeze. Let us use
the phrase \textquotedblleft super Hubble evolution\textquotedblright
~to mean any evolution after $R_k$ has completely frozen once. \emph{In this
super Hubble limit, it is impossible that the amplitude gets
frozen while the phase does not, however, it is possible that the phase
freezes but the amplitude does not}. Writing $v_k = r e^{i \theta}$,
Eq. (\ref{MSE}) implies that
 \begin{equation} \label{theta}
  \theta '' + 2 \left( \frac{r'}{r}\right) \theta' = 0 \,.
 \end{equation}
Notice that, once the phase of $v_k$ (and hence $R_k$) is frozen, it can not
unfreeze. Also, the rate of change of $\theta'$ is directly
proportional to $\theta'$ itself. Hence, the nearer we are to the epoch
of phase freezing for a given mode, the less will the phase get
affected by any background evolution.
Thus, once the mode goes out of the Hubble radius, the phase freezes,
in ordinary scenarios, the amplitude will also freeze.  However, if it
is arranged to unfreeze $R_k$, (by briefly decreasing $z''/z$) even
then only the amplitude will unfreeze and not the phase.  Thus, \textit{if
there is a possibility of any super-Hubble evolution of the mode,
that should lead to only radial trajectory in the complex plane!} (see Fig. \ref{stdR}). Thus, provided that
the phase of the mode function is already frozen, if
such an evolution is sufficiently large, the mode must cross the
origin in the complex plane. This is connected to a cuspy dip in the scalar PPS.

In the rest of the paper, we put forth the conditions that lead to the
nulling in the PPS leading to cuspy dip features. Recall from
Eq.~(\ref{MSE}) that it is the peculiarities in the dynamics of the
quantity $z''/z$ that can lead to super-Hubble evolution.

We seek an evolution of $z''/z$ which is such that the mode function
of $R_k$ for at least some wavenumber, $k$, crosses the origin in the
complex plane. Then simple continuity argument ensures that for
at least one mode, the mode function of $R_k$ would freeze exactly at the origin.

For clarity let us recall that in the simplest case of power law
inflation, $z$ goes as ${\eta}^{1+ \gamma}$ (where $\eta$ is the
conformal time and $\gamma $ is a constant). In such a case, we shall
have, $ \frac{z''}{z~} ~=~ \frac{a''}{a} ~=~ \frac{\gamma (\gamma +1)}
{ \eta ^2} \,.$

\begin{figure}[t]
\begin{center}
\epsfig{file=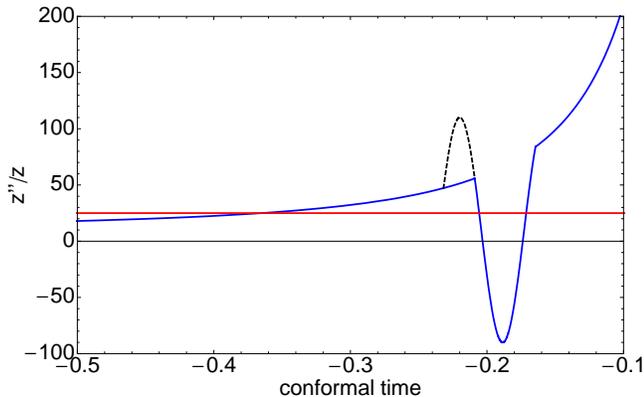,angle=0,width=8.5cm}
\end{center}
\caption[] {\small\sf Typical behaviour of $z''/z$ against conformal time, $\eta$, in any sandwich model (blue curve). The horizontal red line is $k^2$ while the dashed curve is an alternative modified $z''/z$ that may be needed to ensure that the mode of interest becomes super-Hubble (as explained in the text).}
\label{sandwich}
\end{figure}

Note that the first equality implies evolution of scalar modes and
tensor modes are identical in these models, we will use this
observation later. First, it is important to note the monotonic
increasing form of $\frac{z''}{z~}$. In this case, a mode which has
once become super-Hubble ($k^2 < z''/z$) would continue to stay in that
regime and the perturbation $R_k$ will remain frozen. To unfreeze the
mode, it is important for $z''/z$ to decrease, and hence have a
non-monotonic dip like feature (see Fig. \ref{sandwich}).

Hence, we consider a class of approximate models that we refer to as
\textquotedblleft sandwich\textquotedblright ~models. These have three
distinct stages of evolution during inflation. The quantity $z''/z$
follows the monotonic power law inflation evolution in stage I. In
stage II, there is a specific deviation from power law
inflation leading to a desired dip feature in $z''/z$ (Fig. \ref{sandwich}). Stage III
reverts to power law inflation. In what follows, we shall make a
series of statements that will hold good for any such model. We will
illustrate our arguments with the help of (i) Punctuated inflation (PI) model
\cite{PI1,PI2} (see Fig. [\ref{PI_PPS}]), and (ii) Starobinsky-break model
\cite{Staro1992}, which have been well-studied in literature.

The origin of cuspy dips in the power spectrum can now be understood in terms of the following salient features which are summarized below and elaborated afterwards: 
\begin{itemize}
 \item \textbf{Radial trajectory:} Super-Hubble evolution (any evolution after $R_k$ for the mode freezes in stage I) always leads to only radial trajectory in the complex plane.
 \item \textbf{Inward motion:} The super-Hubble evolution involves a radially inward motion (see Eq. (\ref{Delr'})).
 \item \textbf{Amount of super-Hubble evolution:} The amount of super-Hubble evolution in stage II or III is determined by the depth of the dip in $z''/z$ in stage II.
 \item \textbf{Continuity:} If a mode $k_1$ crosses the origin in the complex plane on a radial trajectory, there should exist a mode $k_*$ (with $k_* < k_1$) that ends up right at the origin.
\end{itemize}

Thus, it is clear that one can easily construct sandwich models which will offer cuspy dips in scalar PPS under fairly general conditions.

\begin{figure}
\centerline{\psfig{figure=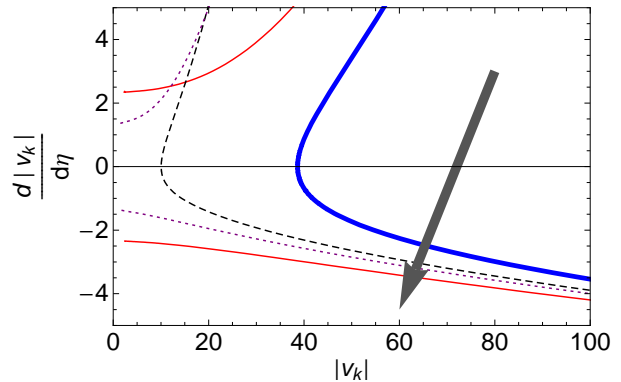, width = 8 cm, angle=0}}
\caption{\small\sf The behavior of the mode in $(r\equiv|v_k|,r'\equiv|v_k|')$ phase space for
a given mode in Starobinsky model for different strengths of the delta
function in $z''/z$. Stage I of the evolution shall make sure that $\theta'$ has already become zero. For low strengths of the delta function, blue(thick) and black(dashed), there is no
origin crossing, while for high strengths, purple(dotted) and red(continuous thin),
$v_k$ passes through origin (leading to a discontinuity in $r'$). Notice that since this is not an autonomous system, the trajectories do cross each other. The thick arrow illustrates that the stage II is expected to take $(|v_k|,|v_k|')$ to regions with sufficiently negative $|v_k|'$.}
\label{starof}
\end{figure}

For a mode whose phase is frozen in stage I,
it is absolutely necessary that, in stage II, $v_k$
should turn back and go radially in if it has to undergo origin
crossing. This is a necessary but insufficient condition for origin crossing as is illustrated for Starobinsky-break model (in which stage II is just a Dirac delta function) in Fig. \ref{starof}. For a given $k$, the dip in
$z''/z$ can be made sufficiently deep, to have origin crossing. Both the quantities $|v_k|$ and $|v_k|'$ are positive in stage I of a sandwich model, so we are in the upper half of Fig. \ref{starof}. The desired model will have a stage II which brings us in the lower half of Fig. \ref{starof} which shows that if $r'$ is sufficiently negative, origin crossing definitely take place. Thus, if
after stage II, both $|v_k|$ and $\frac{d}{d
\eta} |v_k|$ are sufficiently small, origin crossing must take
place.

From Eq. (\ref{MSE}), we can find the equation for the super-Hubble
evolution of amplitude, $r$, of $v_k$. The condition for radial
evolution means that ${\theta}'$ vanishes. The equation for $r$ in
this approximation is,

\begin{equation} \label{r2}
 \frac{r''}{r} + \left[  k^2 - \frac{z''}{z} \right] = 0 \,.
\end{equation}

One can estimate the super-Hubble evolution of $r$
in stage II. We see from Eq. (\ref{r2}) above that the change in
the derivative of the amplitude of $v_k$ is:

\begin{equation} \label{Delr'}
 \Delta {r}' = - \int \left( k^2 - \frac{z''}{z} \right) r ~ d {\eta} \,.
\end{equation}

The obvious conclusion that can be easily drawn from Eq. (\ref{Delr'})
is that for a fixed $k$, unless $z''/z$ is such that $k^2 > z''/z$,
the change in $r'$ will be positive, so that at the end of stage II,
instead of having $v_k$ turn back in the complex plane, we shall have
$v_k$ going radially out at a quicker rate and origin crossing will
most definitely not happen. This means that to cause origin crossing,
for a given mode, we need a sufficiently deep dip such that $k^2 >
z''/z$.

Let us fix $z''/z$ and consider two modes ( $k_1$ and $k_2$ with $k_1 > k_2$)
with sufficiently small $k$
values such that $R_k$
corresponding to both have frozen in stage I ($v_k$  radially outgoing). Stage I is power
law inflation that gives tilted red spectrum so that $P_s(k_2) >
P_s(k_1)$, which means that (since $k_1 > k_2$) $R_{k_2} >
R_{k_1}$. Thus $v_{k_2} > v_{k_1}$. This means that the amplitude
$v_k$ in the complex plane at the end of stage one is smaller for a
mode having larger $k$ value. Also, since $R_k$'s have already frozen,

\begin{equation} \label{v'}
 \frac{v_{2}'}{v_{1}'} = \frac{v_{2}}{v_{1}} > 1
\end{equation}

Thus, the speed with which $|v_k|$ increases in the complex plane at
the end of stage I is also smaller for a mode having larger $k$
value. From Eq. (\ref{Delr'}), we also know that $\Delta ({r}')|_1 >
\Delta ({r}')|_2$. Thus, we conclude the following: for a fixed
$z''/z$, even if origin crossing does not happen for a given $k$ mode
the chances that a larger $k$ mode will undergo origin crossing is
much larger (provided that in stage I, this larger $k$ mode has become super-Hubble i.e. frozen once).

But increasing just the $k$ value for a given model can potentially lead us to modes that have not become super-Hubble in stage I.
In such a case, one could (i) delay the location of dip (in $z''/z$) so that 
it occurs a little later and by that time $v_k$ for that mode
becomes radial, or, (ii) increase $z''/z$ in stage I, (such as the dashed curve in Fig. \ref{sandwich}). In either case, it is easy to see that stage I can be conveniently modified to cause origin crossing in stage III due to a given dip in stage II.

A deep dip in $z''/z$ will cause $z$ to quickly change, as a result,
$|z|$ will actually decrease briefly. Since in an expanding universe,
the scale factor can not decrease, the said evolution can never happen
for tensor modes whose evolution is governed by $a''/a$ (also see \cite{Leach1}). If the scalar
power corresponding to a mode undergoes super Hubble suppression while
the tensor power does not, it is not a surprise that the tensor to
scalar ratio can become greater than one (see Fig. \ref{PI_PPS} and ref.
\cite{PI2}).

Thus, it is important to pay attention to the evolution of the mode functions (of various quantities related to perturbations of interest) in the complex plane. We learn that though the amplitude of $R_k$ can be reawakened once it is frozen, frozen phase can not unfreeze i.e. super-Hubble evolution leads to radial trajectory in the complex plane. In Sandwich models, for modes that have taken up such a radial trajectory in the complex plane, the amount of super-Hubble evolution is determined by the depth of the dip in $z''/z$ in stage II. If a given dip fails to bring the mode to the origin, we can easily modify stage I such that it does so. This \textquotedblleft explains\textquotedblright the origin of (i) nulls in the scalar PPS (ii) tensor power overtaking scalar power, observed in the literature. There is no apriori reason that such zeros in PPS should survive even when higher order corrections are taken into account. Usually, quantum corrections to the PPS are too small compared to the classical result (see e.g. \cite{quantum-recent}), here, since the classical contribution is zero, the truly quantum nature of the primordial perturbations may get revealed.

\noindent {\bf Acknowledgment:} The authors would like to thank
L. Sriramkumar for comments and discussion at various stages of the
work. We would also like to thank Alexei Starobinsky for illuminating
comments and suggestions. GG thanks
Council of Scientific and Industrial Research (CSIR), India, for the
research grant award No. 10-2(5)/2006(ii)-EU II. TS acknowledges
support from the Swarnajayanti Fellowship, DST, India.

\end{document}